# Synthesis, Crystal Structure, Magnetic and Electronic Properties of the Caesium-based Transition Metal Halide $Cs_3Fe_2Br_9$

Fengxia Wei,[a,b] Federico Brivio,[a] Yue Wu,[a] Shijing Sun,[a] Paul D. Bristowe [a]* and Anthony K. Cheetham[a]*

[a]Department of Materials Science and Metallurgy, University of Cambridge, CB3 0FS, UK.
[b]Institute of Materials Research and Engineering, Agency for Science, Technology and Research, 2 Fusionopolis Way, Singapore.
*Corresponding authors: akc30@cam.ac.uk, pdb1000@cam.ac.uk

**The diversity of halide materials related to important solar energy systems such as $CsPbX_3$ (X = Cl, Br, I) is explored by introducing the transition metal element Fe. In particular a new compound, $Cs_3Fe_2Br_9$ (space group $P6_3/mmc$ with a = 7.5427(8) and c = 18.5849(13) Å), has been synthesized and found to contain 0D face-sharing $Fe_2Br_9$ octahedral dimers. Unlike its isomorph, $Cs_3Bi_2I_9$, it is black in color, has a low optical bandgap of 1.65 eV and exhibits antiferromagnetic behavior below $T_N$ = 13 K. Density functional theory calculations shed further light on these properties and also predict that the material should have anisotropic transport characteristics.**

1. Introduction

In the past few years, lead halide perovskites such as $APbI_3$ (A = methylammonium, MA, and cesium) have attracted much attention as photovoltaic materials because of their remarkable photo-conversion efficiency in solar cell devices.[1,2] Due to the toxicity of lead and the intrinsic moisture sensitivity of the lead (II) compounds, a search for environmentally friendly alternatives has been undertaken.[3] Several perovskite-related families have been proposed, such as double perovskites where $Pb^{2+}$ is replaced by isoelectronic $Bi/In/Sb^{3+}$ and a monovalent cation, e.g. $(MA)_2KBiCl_6$, $(MA)_2TlBiBr_6$, $(MA)_2AgBiBr_6$ and the inorganic phases $Cs_2AgBX_6$ (X = Cl, Br and B = Bi, In).[4–11] Another popular candidate family is $A_3M_2I_9$, where A = $K^+$, $Rb^+$, $NH_4^+$, $MA^+$, $Cs^+$ etc, M = $Bi^{3+}$ and $Sb^{3+}$, consisting of either corner- and edge-sharing $MI_6$ octahedral layers  or face-sharing $MI_6$ dimers.[12–16] All of the above systems exhibit very interesting optoelectronic properties.

Transition metals have attracted our attention as a method of tuning the optoelectronic properties. For example, using $Fe^{3+}$ to replace $Bi^{3+}$ can reduce the bandgap: $Cs_2NaFeCl_6$, which adopts a double perovskite architecture (Figure S1, ESI) is red, while its Cl analogues with other trivalent cations show much lighter colours. For instance, $Cs_2NaBiCl_6$ is yellow[17] while the $Cs_2NaLnCl_6$ (Ln = Lanthanide) phases are mostly white.[18] A much darker color is expected for the hypothetical $Cs_2NaFeBr_6$, but our attempts to synthesize this compound yielded black octahedral crystals of composition $Cs_2FeBr_5 \cdot H_2O$ (Figure S2, ESI), crystallizing in space group $Pnma$. This material consists of 0D $FeBr_5O$ octahedral monomers in which the oxygen is part of a water molecule, as in the known $Cs_2FeCl_5 \cdot H_2O$.[19] The dimensionality indicates the degree of connectivity of the octahedra. In this case the octahedra are discrete.

Incorporating Fe into the $A_3Bi_2X_9$ (X = Cl, Br and I) family turns out to have a long history. $Cs_3Fe_2Cl_9$, which is dark red in color, was reported to form two polymorphs: a 2D layered system with $P\bar{3}m1$ symmetry and 0D dimeric system in space group $P6_3/mmc$,[20,21] In the latter, both intradimer and interdimer magnetic interactions are present, and the two competing interactions lead to very interesting magnetic properties. In the present work, we report a new compound, $Cs_3Fe_2Br_9$ (CCDC 1575068), which is isostructural with $Cs_3Bi_2I_9$ (red)[13] and $MA_3Bi_2I_9$ (red),[14] yet is black in color. Its variable temperature behavior, thermal stability, optical and magnetic properties are investigated in combination with density functional theory (DFT) calculations.

2. Experimental and computational methods and results

2.1 Synthesis

A two-step synthesis method was used, involving both hydrothermal and room temperature crystallization. 2 mmol CsBr (99.9%, Sigma Aldrich), 1 mmol $FeCl_3 \cdot 6H_2O$ (>99%, Sigma Aldrich) together with 1.5 ml HBr acid (47 wt%) were placed in a 23 ml stainless steel Parr autoclave and heated at 160°C for 3 days. Intermediate products of brown needle shaped crystals of $CsFeBr_4$ (Figure S3, ESI) were formed. The Teflon autoclave was then left in the fume hood at room temperature (>15°C) and black crystals formed after one week. The following chemical reactions take place during the synthesis:

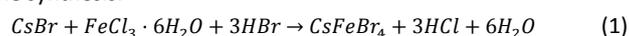
$$CsBr + FeCl_3 \cdot 6H_2O + 3HBr \rightarrow CsFeBr_4 + 3HCl + 6H_2O \qquad (1)$$

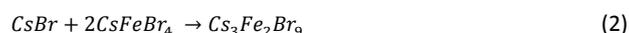
$$CsBr + 2CsFeBr_4 \rightarrow Cs_3Fe_2Br_9 \qquad (2)$$

During the hydrothermal process, reaction (1) dominates and almost no black $Cs_3Fe_2Br_9$ is formed. Even using exact stoichiometric ratios of the starting reagents does not result in the target material. However, black octahedral crystals of $Cs_3Fe_2Br_9$, ~ 0.5 mm in size, can be collected after standing at room temperature for 3 weeks. The sample is soluble in most polar solvents, including water, ethanol and acetone.

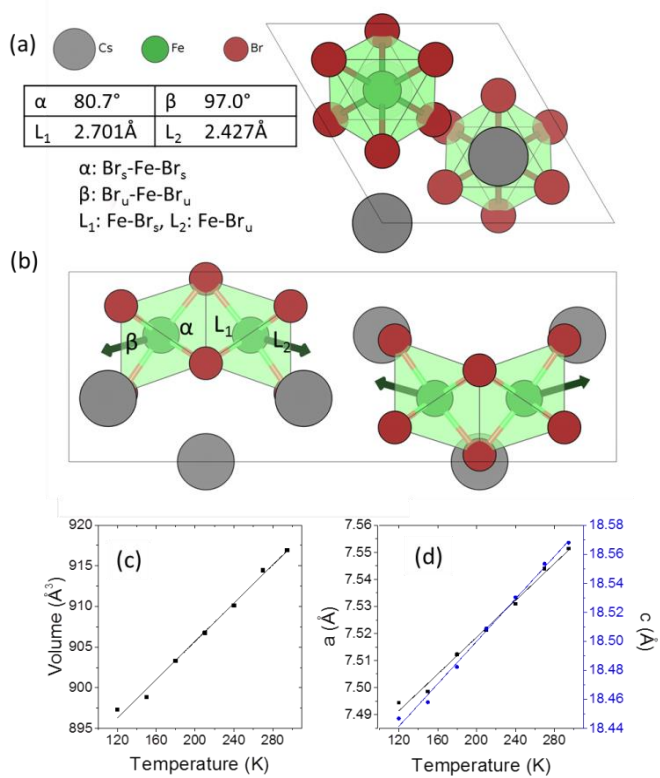

**Figure 1.** (a) Crystal structure of $Cs_3Fe_2Br_9$ viewed along the *c*-axis, (b) view along the *b*-axis showing of the $Fe_2Br_9$ dimers. The angles and bond lengths illustrate the distortion of the octahedra. The subscripts s and u indicate shared and unshared Br anions respectively and the arrows indicate the direction of Coulombic repulsion between cations, (c) cell volume and (d) lattice parameters as a function of temperature measured using single crystal X-ray diffraction.

### 2.2 Crystallographic studies

$Cs_3Fe_2Br_9$ crystallizes in the hexagonal space group $P6_3/mmc$ (a = 7.5427(8) and c = 18.5849(13) Å). It consists of face–sharing $Fe_2Br_9$ octahedral dimers with Cs serving as bridging atoms between the dimers (Figure 1a, b). The octahedra are slightly distorted, with two sets of Fe–Br bonds (2.427(1) Å and 2.701(2) Å) and distorted Br–Fe–Br angles (80.76(6)°, 90.55(3)° and 97.01(7)°), compared to the nominal octahedral angle of 90°. Due to the Coulombic repulsive force between the cations within the dimer (Fe-Fe distance = 3.585(3) Å), the $Fe^{3+}$ ions are displaced outwards with respect to the shared face. Therefore, the smallest octahedral angles and longer Fe–Br bonds are found with the shared $Br^-$ ions (Figure 1) and the largest angles and shorter Fe–Br distances are from the unshared ones. According to the interatomic distances, the bond strengths between $Fe^{3+}$ and unshared $Br^-$ are stronger than those with shared $Br^-$ ions. Moreover, the angular distortion of the $Br_{shared}$–Fe–$Br_{unshared}$ angle is minor (90.55°). The shortest distance between Cs and Br is 3.762(1) Å.

Variable temperature single crystal diffraction suggests no phase transition down to 120 K. The thermal expansion coefficients are approximately linear with $α_a$ = 45.3 $MK^{-1}$, $α_c$ = 39.6 $MK^{-1}$, giving $α_v$ = 131.2 $MK^{-1}$. The repulsion between the $Fe^{3+}$ ions in the dimeric unit decreases upon cooling, as shown by the less distorted octahedral dimer and reduced interatomic Fe…Fe distances (Figure S4, ESI). As a result, negative expansion occurs for the shorter bonds and positive thermal expansion is found for the longer bonds. A similar phenomenon is observed for the octahedral angles: on cooling, the smaller angles tend to increase, while the larger angles decrease.

### 2.3 Thermal analysis

Thermal stability was investigated using an SDT (simultaneous differential scanning calorimetry (DSC) - thermogravimetric analysis (TGA)) Q600 instrument. Powder samples were heated from room temperature to 1123 K at 10 K/min under an air flow of 100 ml/min. $Cs_3Fe_2Br_9$ is stable until 537.5 K and then experiences a two-step decomposition process. When the sample is heated, moisture and residual HBr at the particle surfaces start to evaporate, resulting in a small weight loss (~3.6%) at the beginning of the curve. For comparison, the thermal stability of its bismuth analogues Cs$_3$Bi$_2$I$_9$ and MA$_3$Bi$_2$I$_9$ were also measured; the former decomposed at 636.4 K, while the latter was stable until 529.3 K (Figure S5 and S6, ESI).

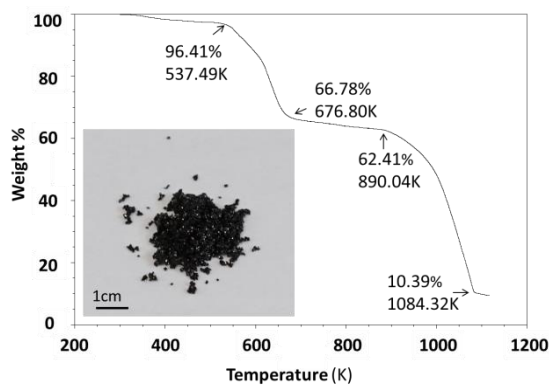

**Figure 2.** Thermogravimetric analysis curve; the inset shows a photo of the sample (small crystals).

### 2.4 Optical characterization

The optical bandgap was measured on a PerkinElmer Lambda 750 UV-Visible spectrometer in the absorption mode with a 2nm slit width. The scan interval was 1 nm and the scan range was between 500 and 1100nm. The absorption edge is observed at ~ 800nm (i.e. 1.55 eV). In accordance with our DFT calculation (see below), we deduced a direct optical bandgap of ~ 1.65 eV from the Tauc plot derived from the reflectance spectrum (Figure 3). Note that the analogous A$_3$Bi$_2$I$_9$ phases (A = Cs and MA) were reported to have indirect bandgaps which are larger in the range 1.9 eV to 2.2 eV.[22,23]

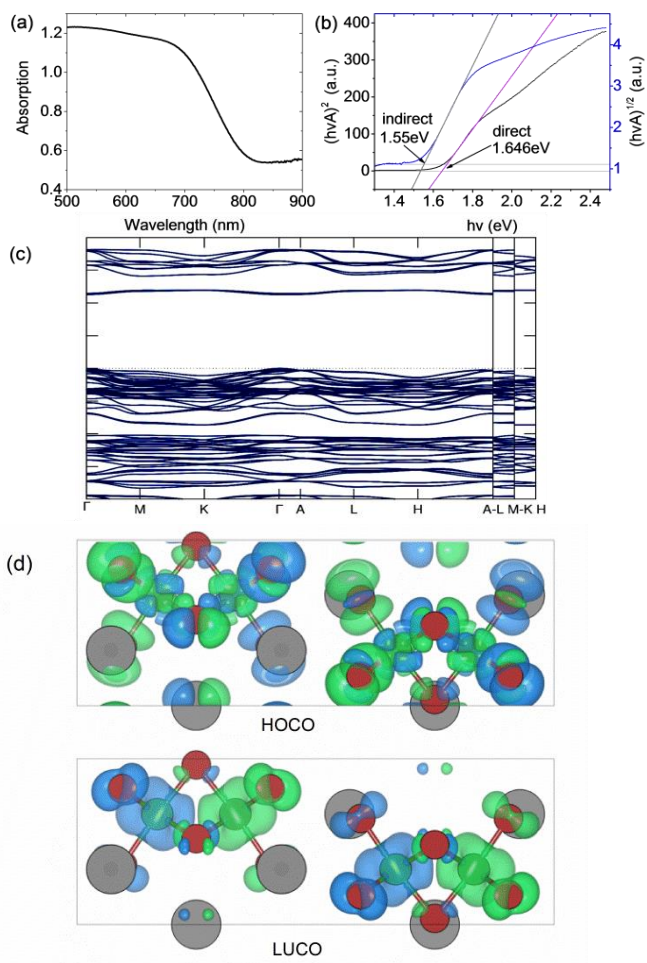

**Figure 3.** (a) Absorption spectrum and (b) Tauc plot for indirect and direct bandgaps. (c) Band structure (non-magnetic case) calculated using the HSE06 exchange-correlation functional. (d) Charge density isosurfaces (antiferromagnetic case) calculated using the PBEsol exchange-correlation functional and viewed along the *b*-axis. The top and bottom panels show the Highest Occupied Crystal Orbital (HOCO) and Lowest Unoccupied Crystal Orbital (LUCO) respectively. The charge is displayed using a threshold of 0.001 e/Bohr$^3$. The different spin channels are shown in blue and green. Atom colors are the same as in Figure 1.

## 2.5 Density functional calculations

The DFT calculations were performed using the projector augmented wave (PAW) method as implemented in VASP.[24] The experimental structure obtained at room temperature was fully optimized using the PBEsol exchange-correlation functional[25] which reduced the atomic forces below 1 meV/Å at effectively zero Kelvin (see ESI for further computational details). The resulting atomic positions are given in Table S1. The presence of Fe in the material suggests that it could exhibit magnetic ordering due to unpaired 3*d* electrons. To examine this possibility, spin-polarized calculations were performed on the optimized structure in the ferromagnetic (FM) state and three possible antiferromagnetic (AFM) states. It was found that one of the AFM states in which neighboring Fe atoms have opposite spin orientation is significantly lower in energy than either the FM or non-magnetic states, by 80 meV/f.u. and 335 meV/f.u. respectively (see Table S2 for details). The calculations therefore predict that at very low temperatures $Cs_3Fe_2Br_9$ prefers to be antiferromagnetic. The calculated magnetic moment on each Fe atom is 3.38 $\mu_B$. This value is lower than the value of 5.79 $\mu_B$ obtained from analysis of the magnetic susceptibility data in the higher temperature paramagnetic region (see below). There are several reasons for this, including the well-known reduction in spin in magnetically ordered structures due to covalency. For example, neutron scattering measurements on $FeCl_3$ show that the spin is reduced to 4.7(3) $\mu_B$ in the antiferromagnetic phase.[26] Figure 3 shows charge density isosurfaces corresponding to the HOCO and LUCO for the lowest energy AFM state.

In order to determine an accurate band structure for $Cs_3Fe_2Br_9$ the HSE06 hybrid exchange-correlation functional was used.[27] The calculation was performed on the non-magnetic state to contain the cost of the calculation and because previous work has indicated that while HSE06 returns a better description of the band gap, it may not be adequate for magnetic properties.[28] The material is found to have a 2.254 eV direct band gap which occurs at the Γ point with a relatively flat band structure (Figure 3). At the band edge it is possible to calculate the effective masses in the parabolic approximation (Table 1). The values indicate a high anisotropy with reduced transport along the *c*–direction (Γ→A). The Fe atoms have been described with $3p^63d^74s^1$ as valence electrons, while other core states have been substituted by the pseudopotential. The valence band maximum (VBM) contains Fe 3d and Br 4p states, whereas the conduction band minimum (CBM) contains mostly Fe 3d, Fe 4s and Br 4p states.

**Table 1.** Calculated effective masses (relative to the rest mass $m_0$).

|  | Γ→M | Γ→K | Γ→A |
|---|---|---|---|
| $m_h^*/m_o$ | -0.11 | -0.06 | -1.02 |
| $m_e^*/m_o$ | 0.25 | 0.16 | 13.87 |

## 2.6 Magnetic measurements

Magnetic susceptibility measurements, $\chi(T) = M(T)/H$, were conducted using a Quantum Design Magnetic Properties Measurement System (MPMS3) with a superconducting interference device (SQUID) magnetometer. Measurements were made after cooling in zero field (ZFC) and in a measuring field (FC) of $\mu_0 H$ = 0.01 T over the temperature range 2 ≤ T ≤ 300 K. $Cs_3Fe_2Br_9$ shows antiferromagnetic behavior with a Néel temperature $T_N$ = 13 K (Figure 4), higher than that of analogous $Cs_3Fe_2Cl_9$ which also exhibits an antiferromagnetic long range order at $T_N$ = 5.3 K.[29] The results are in good agreement with the DFT calculations.

Appling the Curie-Weiss law in the paramagnetic region (from 50 K to 300 K), a negative Weiss constant of -36.10(3) K was obtained, as expected for an antiferromagnetic compound (Figure S7, ESI) and the calculated effective magnetic moment $\mu_{eff}$ was 5.79(4) $\mu_B$. In order to obtain a comprehensive χ fitting from 2 K to 300 K, a weakly coupled dimer model was applied.[30] This system contains $Fe^{3+}$ ions as dimeric units $Fe_2Br_9$, with three distinct Fe–Fe distances (3.585(3), 7.179(2) and 7.543(1) Å) corresponding to intradimer (J) and interdimer ($J_c + J_p$) interactions, respectively (see ESI for fitting formula). The dominant intradimer exchange yields J = -8.2 K, while the weak interdimer interactions between the $Fe^{3+}$ of neighbouring dimers are given as $J_p + J_c$ = -3.4 K. The intradimer interactions are weaker than those in the Cr counterpart, $Cs_3Cr_2Br_9$, which was reported to have J = -10.3 K, $J_p + J_c$ = -1.1 K; on the other hand, the interdimer interactions of $Cs_3Fe_2Br_9$ are stronger.[30] In the case of $Cs_3Fe_2Cl_9$, both intra- and interdimer interactions are smaller (J = -2.4 K and $J_p + J_c$ = -1.2 K for single crystals).[29]

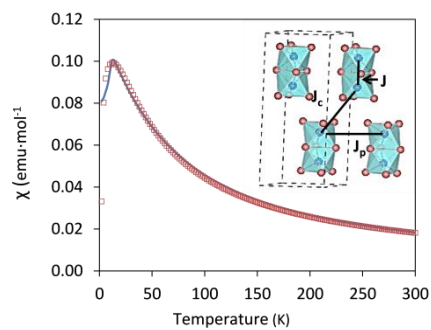

**Figure 4.** Magnetic susceptibility as a function of temperature, blue: experimental, red square: fitted curve using dimer model. The inset illustrates the definition of J, $J_p$ and $J_c$. Cs atoms are not included for clarity.

## Conclusions

We have synthesized black crystals of $Cs_3Fe_2Br_9$ and determined its crystal structure. $Cs_3Fe_2Br_9$ crystallizes in the hexagonal space group $P6_3/mmc$ and the structure contains $Fe_2Br_9$ face-sharing octahedral dimers. The Fe-Br bond strengths differ between the shared and unshared faces, and the distorted octahedra tend to become more symmetrical upon cooling due to the reduced cation–cation Coulombic repulsion. The compound is thermally stable up to 537.5 K, and has an optical bandgap of 1.65 eV. DFT calculations indicate that the band gap is direct and also predict reduced transport along the *c*-direction. Magnetic susceptibility measurements show antiferromagnetic behavior, with $T_N$ = 13 K, and can be fitted with a weakly coupled dimer model. The spin polarized DFT calculations agree with this behavior at low temperatures and predict which antiferromagnetic configuration is preferred.

## Conflicts of interest

There are no conflicts to declare.

## Acknowledgements

F. Wei is a holder of an A*STAR international fellowship granted by the Agency for Science, Technology and Research, Singapore. A. K. Cheetham and Y. Wu thank the Ras al Khaimah Center for Advanced Materials for financial support. The calculations were performed at the Cambridge HPCS and the UK National Supercomputing Service, ARCHER. Access to the latter was obtained via the Materials Chemistry Consortium and funded by EPSRC under Grant Number EP/L000202/1.

## Notes and references

**Synthesis, Crystal Structure, and Magnetic and Electronic Properties of the Caesium-based Transition Metal Halide Cs$_3$Fe$_2$Br$_9$**

Fengxia Wei,[a,b] Federico Brivio,[a] Yue Wu,[a] Shijing Sun,[a] Paul D. Bristowe [a*] and Anthony K. Cheetham [a]*

[a]Department of Materials Science and Metallurgy, University of Cambridge, CB3 0FS, UK
[b]Institute of Materials Research and Engineering, Agency for Science, Technology and Research, 2 Fusionopolis Way, Singapore.

*Table of contents*

**Details of methods** including single crystal X-ray diffraction, DFT calculations, and magnetic susceptibility fitting formula.

**Figure S1**. Crystal structure of Cs$_2$NaFeCl$_6$: purple - FeCl$_6$ octahedra, brown - NaCl$_6$ octahedra; Cs is in the cavity of 3D octahedral framework. Space group: $Fm\bar{3}m$, cubic, a = 10.3403(1) Å. Crystal colour: red. CCDC No. 1575067.

**Figure S2**. Structure of Cs$_2$FeBr$_5$·H$_2$O: the distorted FeBr$_5$O is shown in brown, oxygen is red with H attached. a = 14.6652(3) Å, b = 10.4144(2) Å and c = 7.5110(2) Å, space group *Pnma*. CCDC No. 1575069.

**Figure S3**. Crystal structure of CsFeBr$_4$: Purple - FeBr$_4$ tetrahedra. Space group: Pnma, orthorhombic, a = 12.0861(5) Å, b = 7.4271(3) Å, c = 9.8187(4) Å.  Crystal colour: brown. CCDC No. 1575070.

**Figure S4**. Interatomic distances and octahedral angles - variations as a function of temperature.

**Figure S5**. Thermogravimetric analysis curve for Cs$_3$Bi$_2$I$_9$.

**Figure S6**. Thermogravimetric analysis curve for MA$_3$Bi$_2$I$_9$.

**Figure S7**. Curie-Weiss fitting for the paramagnetic region of Cs$_3$Fe$_2$Br$_9$, (a) for χ and (b) for 1/χ to obtain the Weiss constant (fitted from 50 K to 300 K).

**Table S1.** DFT calculated atomic structure (fractional coordinates) of Cs$_3$Fe$_2$Br$_9$.

**Table S2**. DFT calculated energy differences between different magnetically ordered states of Cs$_3$Fe$_2$Br$_9$ relative to the most stable configuration (i.e. AFM-3). U: spin up, D: spin down.

**Details of methods**

Experimental

Crystal structure determination was carried out using an Oxford Gemini E Ultra diffractometer, Mo Kα radiation (λ = 0.71073 Å), equipped with an Eos CCD detector. Variable temperature data were collected from 300 K to 120 K using a Cryo system under liquid nitrogen flow with 30 K steps; the crystal stayed under nitrogen flow for 10 mins at each temperature, allowing sufficient time for cooling. Data collection and reduction were conducted using CrysAliPro. An empirical absorption correction was applied with the Olex2 platform, and the structure was solved using ShelXS[1] and refined with ShelXL.[2]

Computational

A plane-wave cut-off of 600 eV and a 6x6x1 k-point grid centred on the Γ point was used during geometry optimisation of $Cs_3Fe_2Br_9$. For the band structure calculation using HSE06 the plane wave cut-off was reduced to 500 eV. The optimised atomic structure is given in Table S1.

Three possible anti-ferromagnetic (AFM) configurations labelled UU-DD (AFM-1), UD-DU (AFM-2) and UD-UD (AFM-3) were considered for the four Fe atoms in a unit cell, where U denotes spin up and D denotes spin down. The energies of these configurations are compared to the ferromagnetic (FM) and non-magnetic (NM) cases in Table S2. The most stable configuration is the one where next neighbour Fe atoms have opposite spin orientation.

Formula for fitting magnetic susceptibility

The magnetic susceptibility can be expressed by intradimer J and interdimer interactions $J_p + J_c$:

$$\chi(T) = \chi_o(T)/[1 - 3(J_p + J_c) \cdot \frac{\chi_o(T)}{(g^2 \mu_B^2 N_A)}]$$

and

$$\chi_o(T) = \left[\frac{2 N_A g^2 \mu_B^2}{kTZ}\right] \left[\exp\left(\frac{J}{kT}\right) + 5 \exp\left(\frac{3J}{kT}\right) + 14 \exp\left(\frac{6J}{kT}\right)\right]$$

with

$$Z = 1 + 3 \exp\left(\frac{J}{kT}\right) + 5 \exp\left(\frac{3J}{kT}\right) + 7\exp\left(\frac{6J}{KT}\right)$$

where $N_A$ is Avogadro's number, $\mu_B$ is the Bohr magneton, g is a dimensionless magnetic moment or g-value, and k is the Boltzmann constant.

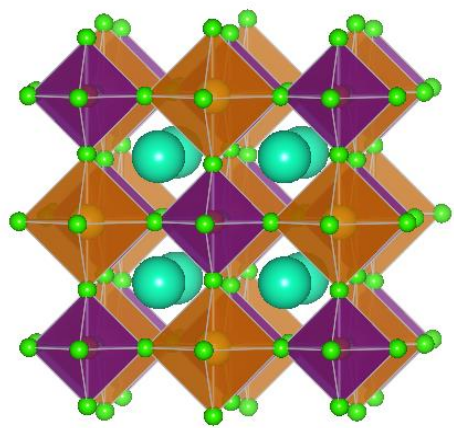

Figure S1. Crystal structure of $Cs_2NaFeCl_6$: purple - $FeCl_6$ octahedra, brown - $NaCl_6$ octahedra; Cs is in the cavity of 3D octahedral framework. Space group: $Fm\bar{3}m$, cubic, a = 10.3403(1) Å. Crystal colour: red. CCDC No. 1575067.

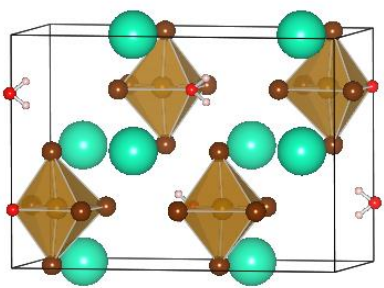

Figure S2. Structure of $Cs_2FeBr_5·H_2O$: the distorted $FeBr_5O$ is shown in brown, oxygen is red with H attached. a = 14.6652(3) Å, b = 10.4144(2) Å and c = 7.5110(2) Å, space group *Pnma*. CCDC No. 1575069.

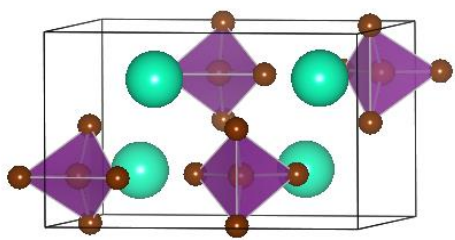

Figure S3. Crystal structure of $CsFeBr_4$: Purple - $FeBr_4$ tetrahedra. Space group: Pnma, orthorhombic, a = 12.0861(5) Å, b = 7.4271(3) Å, c = 9.8187(4) Å. Crystal colour: brown. CCDC No. 1575070.

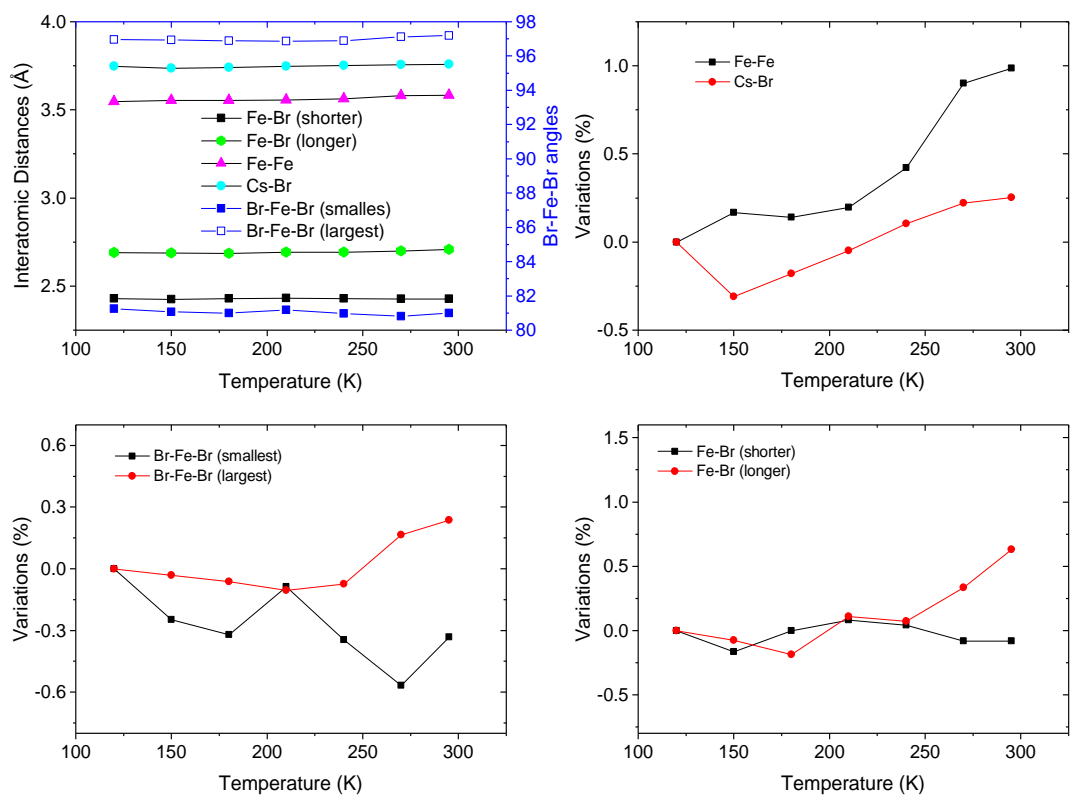

Figure S4. Interatomic distances and octahedral angles - variations as a function of temperature.

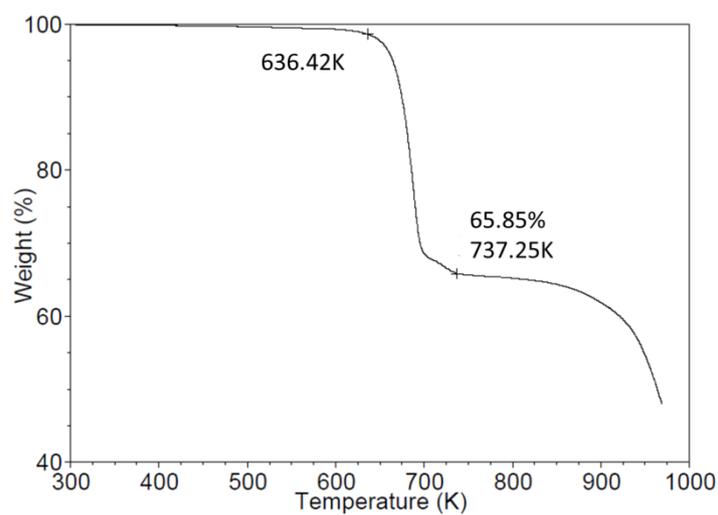

Figure S5. Thermogravimetric analysis curve for Cs$_3$Bi$_2$I$_9$.

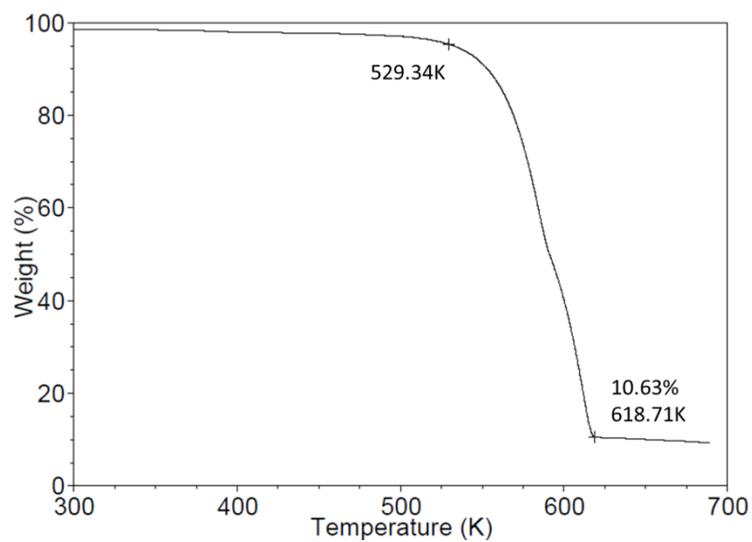

Figure S6. Thermogravimetric analysis curve for MA$_3$Bi$_2$I$_9$.

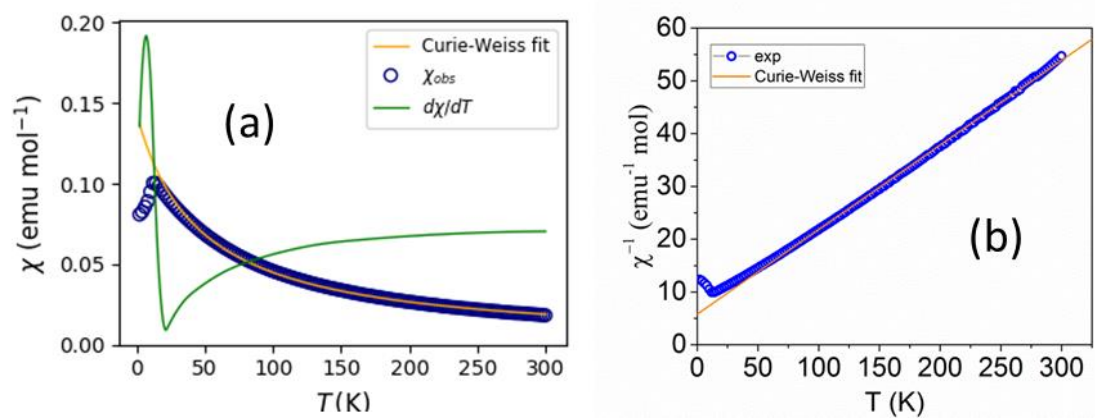

Figure S7. Curie-Weiss fitting for the paramagnetic region of Cs$_3$Fe$_2$Br$_9$, (a) for χ and (b) for 1/χ to get the Weiss constant (fitted from 50 K to 300 K).

Table S1. DFT calculated atomic positions of $Cs_3Fe_2Br_9$. Lattice parameters: a = b = 7.4201Å, c = 18.3609 Å, α = β = 90°, γ = 120°, vol = 875.467 Å$^3$, density = 4.6642 g/cm$^3$.

| No. | label | Fractional coordinates | | | Orthogonal Coordinates | | |
|---|---|---|---|---|---|---|---|
| | | x | y | z | $x_o$ [Å] | $y_o$ [Å] | $z_o$ [Å] |
| 1 | Br1 | 0.036950 | 0.518475 | 0.75 | -0.2374 | 3.71 | -13.7707 |
| 2 | Br2 | 0.963050 | 0.481525 | 0.25 | -6.1885 | 0.0000 | -4.5902 |
| 3 | Br3 | 0.481525 | 0.518476 | 0.75 | -3.0943 | 2.0607 | -13.7707 |
| 4 | Br4 | 0.518475 | 0.481524 | 0.25 | -3.3317 | 1.6494 | -4.5902 |
| 5 | Br5 | 0.481524 | 0.963050 | 0.75 | -3.0943 | 5.3594 | -13.7707 |
| 6 | Br6 | 0.518476 | 0.036950 | 0.25 | -3.3317 | -1.6494 | -4.5902 |
| 7 | Br7 | 0.180243 | 0.360485 | 0.595179 | -1.1582 | 2.0061 | -10.928 |
| 8 | Br8 | 0.819757 | 0.639515 | 0.404821 | -5.2677 | 1.7039 | -7.4329 |
| 9 | Br9 | 0.639515 | 0.819758 | 0.595179 | -4.1095 | 3.71 | -10.928 |
| 10 | Br10 | 0.360485 | 0.180242 | 0.404821 | -2.3165 | -0.0000 | -7.4329 |
| 11 | Br11 | 0.180242 | 0.819757 | 0.595179 | -1.1582 | 5.4139 | -10.928 |
| 12 | Br12 | 0.819758 | 0.180243 | 0.404821 | -5.2677 | -1.7039 | -7.4329 |
| 13 | Br13 | 0.819757 | 0.639515 | 0.095179 | -5.2677 | 1.7039 | -1.7476 |
| 14 | Br14 | 0.180243 | 0.360485 | 0.904821 | -1.1582 | 2.0061 | -16.6133 |
| 15 | Br15 | 0.360485 | 0.180242 | 0.095179 | -2.3165 | -0.0000 | -1.7476 |
| 16 | Br16 | 0.639515 | 0.819758 | 0.904821 | -4.1095 | 3.71 | -16.6133 |
| 17 | Br17 | 0.819758 | 0.180243 | 0.095179 | -5.2677 | -1.7039 | -1.7476 |
| 18 | Br18 | 0.180242 | 0.819757 | 0.904821 | -1.1582 | 5.4139 | -16.6133 |
| 19 | Cs1 | 0.000000 | 0.000000 | 0.75 | 0.0000 | 0 | -13.7707 |
| 20 | Cs2 | 0.000000 | 0.000000 | 0.25 | 0.0000 | 0.0000 | -4.5902 |
| 21 | Cs3 | 0.666667 | 0.333333 | 0.571456 | -4.2840 | 0 | -10.4925 |
| 22 | Cs4 | 0.333333 | 0.666667 | 0.428544 | -2.1420 | 3.7100 | -7.8685 |
| 23 | Cs5 | 0.333333 | 0.666667 | 0.071456 | -2.1420 | 3.7100 | -1.312 |
| 24 | Cs6 | 0.666667 | 0.333333 | 0.928544 | -4.2840 | 0 | -17.0489 |
| 25 | Fe1 | 0.333333 | 0.666667 | 0.668257 | -2.1420 | 3.71 | -12.2698 |
| 26 | Fe2 | 0.666667 | 0.333333 | 0.331743 | -4.2840 | -0.0000 | -6.0911 |
| 27 | Fe3 | 0.666667 | 0.333333 | 0.168257 | -4.2840 | -0.0000 | -3.0894 |
| 28 | Fe4 | 0.333333 | 0.666667 | 0.831743 | -2.1420 | 3.71 | -15.271 |

Table S2. DFT calculated energy differences between different magnetically ordered states of $Cs_3Fe_2Br_9$ relative to the most stable configuration (i.e. AFM-3). U: spin up, D: spin down.

| Configuration | Energy (eV / unit cell) | Energy difference (meV / unit cell) | Energy difference (meV/f.u.) |
|---|---|---|---|
| NM | -101.94543 | 671 | 335.5 |
| FM | -102.4561 | 160 | 80 |
| AFM-1(UU-DD) | -102.60318 | 13 | 6.5 |
| AFM-2 (UD-DU) | -102.45387 | 162 | 81 |
| AFM-3 (UD-UD) | -102.6164 | 0 | 0 |